# Closed π–electron Network in Large Polyhedral Multi-shell Carbon Nanoparticles [§]

A. I. Shames, [1*] E. A. Katz, [2,3] A. M. Panich, [1] D. Mogilyansky, [4] E. Mogilko, [5] J. Grinblat, [6] V. P. Belousov, [7] I. M. Belousova, [7] and A. N. Ponomarev [8]

[1] *Department of Physics, Ben-Gurion University of the Negev, Be'er-Sheva 84105, Israel*

[2] *Department of Solar Energy and Environmental Physics, J. Blaustein Institutes for Desert Research, Ben-Gurion University of the Negev, Sede Boqer 84990, Israel*

[3] *Ilse-Katz Center for Meso- and Nanoscale Science and Technology, Ben-Gurion University of the Negev, Be'er-Sheva 84105, Israel*

[4] *Institutes for Applied Research, Ben-Gurion University of the Negev, Be'er-Sheva 84105, Israel*

[5] *Department of Physics and Center for Superconductivity, Bar-Ilan University, Ramat Gan 52900, Israel*

[6] *Department of Chemistry, Bar-Ilan University Ramat Gan 52900, Israel*

[7] *Institute for Laser Physics, SIC Vavilov SOI, St. Petersburg 199034, Russia*

[8] *JSC Astrin-Holding, St. Petersburg 19809, Russia*

---

[*] Corresponding Author: Fax: (+972)-8-647 2904; E-mail: sham@bgu.ac.il

[§] **Essential explanation**

This manuscript was accepted for the publication in Phys. Rev. B on 22 December 2008 and scheduled for *Phys. Rev. B*, **79** (1) (2009) issue. Unfortunately during the final preparation of the manuscript another article on the same subject submitted by the same group of authors as a post-conference publication appeared in print in *Diamond & Related Materials,* **18** (2009) 505-510 and may be found here: http://dx.doi.org/10.1016/j.diamond.2008.10.056

Since some overlap between these two articles is quite evident, the manuscript accepted to PRB was (by our approbation) suspended by PRB editors and most likely will not see the light of the day on PRB pages. However it is worth mentioning that the article published in DRM was prepared as a typical brief post-conference paper reporting the main results and first conclusions of the on-going study. On the other hand the text of this manuscript contains new results and analysis that have not been published in DRM since these results and analysis were done after the submission to DRM. We suppose that these results are very important for the nanocarbon society.

**Abstract**

High Resolution Transmission Electron Microscopy (HRTEM), X-ray Diffraction (XRD) and Raman spectroscopy reveal a polyhedral multi-shell fullerene-like structure of astralen carbon nanoparticles. The polyhedra consist of large flat graphitic faces connected by defective edge regions with presumably pentagon-like structure. The faces comprise a stacking of 20-50 planar graphene sheets with inter-sheet distance of ~ 0.340 nm. Average sizes of the particles and their flat faces are ~ 40 nm and ~ 15 nm, respectively. The astralen particles are suggested to have defect-free $sp^2$ flat faces and all defects condense at their polyhedral edges. Electron Paramagnetic Resonance (EPR) spectra of polycrystalline astralen samples reveal two components: a very broad signal with $\Delta H_{pp} > 1$ T and an asymmetric narrow one centered close to $g = 2.00$. The latter consists of two overlapping Lorentzian lines. All spectral components are independent of ambient pressure. The intensities of all EPR signals show no changes on decreasing temperature from $T = 300$ K down to 4 K demonstrating the Pauli paramagnetism. Electron spin-lattice relaxation times $T_{1e}$ remain very short within the same temperature range. Temperature dependent $^{13}$C Nuclear Magnetic Resonance (NMR) measurements yield nuclear spin-lattice relaxation times $T_{1n} \sim T^{-0.612}$. The exponent in the $T_{1n}(T)$-dependence for astralen falls between the metallic behavior, $T_{1n} \sim T^{-1}$ (Korringa relation), and the semiconductor behavior, $T_{1n} \sim T^{-0.5}$. All unusual magnetic resonance features of astralen are attributed to delocalized charge carriers which amount considerably exceeds that of spins localized in defects on multi-shell polyhedra edges.

# 1. INTRODUCTION

The discovery[1] of *buckminsterfullerene*, $C_{60}$, a new variety of carbon, has generated enormous interest in many areas of physics, chemistry and material science. Furthermore, it turns out that $C_{60}$ is only the most abundant member of an entire class of fully conjugated, all-carbon molecules - the fullerenes ($C_{20}$, $C_{24}$, $C_{26}$... $C_{60}$... $C_{70}$, $C_{72}$, $C_{74}$ ... carbon nanotubes). It is already clear from the vast and increasing literature on fullerene-based materials that much complex and fascinating physics lies behind the geometrical simplicity of these structures. Now, the term *fullerenes* is used for various all-carbon closed-cage molecules in which π-conjugated carbon hexagon rings are condensed to form a closed surface with the participation of pentagon rings. In other words, fullerenes can be considered as small pieces of graphite which acquired curvature due to the presence of pentagonal rings in the hexagonal graphitic sheet. Topological analysis on the basis of Euler's theorem for convex polyhedra[2] shows that any fullerene molecule must have 12 pentagonal faces and the number of hexagonal faces is arbitrary.

Recently, much attention has turned towards novel fullerene-related materials, carbon *nanocapsules* or carbon *onions* that consist of concentric fullerene-like shells. The formation of carbon onions was first reported by Ugarte.[3] He observed that carbon soot particles and tubular graphitic structure are transformed into quasi-spherical carbon onions by intense electron-beam irradiation in a transmission electron microscope. The structure of spherical onions cannot be described in terms of perfect fullerene-like shells made of pentagonal and hexagonal carbon rings only and requires introduction of a large number of defects, such as heptagonal-pentagonal pairs.[4] As a result, π-electrons in spherical onions are localized in very small domains of $sp^2$ graphitic sheets, and do not act as conduction electrons.[5] On the contrary, polyhedral onions have an ordered graphitic structure, presumably with defect-free *sp*$^2$ flat faces and a definite number of pentagon-like defects condenses at the polyhedral cusps.[6] Accordingly, delocalized π-electrons in polyhedral onions can act as conduction electrons. This effect should be enhanced as sizes of polyhedra increase. Nowadays, knowledge on the correlation between structure and electronic properties of individual carbon onions is very limited. To the best of our knowledge, such comprehensive studies were performed only for spherical and small polyhedral onions produced by high temperature (HT) vacuum annealing of nanodiamond particles.[5, 7–9]

For the present study we adopted large polyhedral multi-layer carbon particles produced by arc discharge of graphite,[10] named by their discoverers as astralens. Although delocalization of π-electrons might be considered as a prerequisite for the recently reported applications of astralen in nonlinear optics,[11-13] their polyhedral fullerene-like structure has not been confirmed directly so far. We provided such a confirmation using High Resolution Transmission Electron Microscopy (HRTEM), X-ray Diffraction (XRD) and Raman spectroscopy studies. By means of Electron Paramagnetic Resonance (EPR) and Nuclear Magnetic Resonance (NMR) spectroscopies we investigated electronic properties of astralen and demonstrated their qualitative difference from those of the carbon onions obtained from HT treated (HTT) nanodiamonds.[5,8,9] The most unusual feature of the EPR properties of astralens is the Pauli-type behavior all EPR signals in the wide temperature range. To the best of our knowledge, such behavior has never been reported for nanocarbon samples. These results verify a hypothesis that each astralen nanoparticle constitutes a closed network of delocalized π-electrons. The $^{13}$C NMR findings support this suggestion.

## 2. EXPERIMENT

Astralen particles were produced by thermal vaporization of graphite anode by arc discharge using special conditions of vaporization, extraction and subsequent treatment of cathodic deposit.[10] Reference graphite powder sample was obtained by pulverizing Alpha Aesar graphite rod (99 %).

HRTEM images were taken with JEOL 2011 microscope having an acceleration voltage of 200 kV. XRD data were collected on Philips1050/70 powder diffractometer using $CuK_\alpha$-radiation and operating at 40 kV/30 mA. The XRD diffraction patterns were then treated using the FULLPROF program in order to derive positions and widths of X-ray reflections. Raman spectra were recorded with a Jobin-Yvon LabRam HR 800 micro-Raman system equipped with a liquid-$N_2$-cooled detector. The excitation wavelength was He-Ne laser supplied with the JY Raman spectrometer (633 nm). The measurements were taken with the 600 g mm$^{-1}$ grating and a microscope confocal hole setting of 100 μm, giving a resolution of 4–8 cm$^{-1}$. Laser intensity on the sample was 3 mW; the spot size is about 20 μm$^2$. EPR measurements within the temperature range 4 (±0.1) K ≤ $T$ ≤ 300 (±0.5) K were carried out using a Bruker EMX220 X-band (ν ~ 9.4 GHz) spectrometer equipped with an Oxford Instrument ESR900 cryostat and an Agilent 53150A frequency counter. Several mg of astralen (as fabricated) and graphite reference powders were placed in 1 mm i.d. capillary tubes centered into the rectangular EPR cavity aiming to reduce strong non-resonant microwave absorption observed. EPR spectra were recorded for samples in open (to ambient oxygen) and sealed tubes. The latter was done at the vacuum level better than 2 × 10$^{-5}$ mbar. Temperature dependences of the following EPR spectra parameters were analyzed: resonance field $H_r$, peak-to-peak line width $\Delta H_{pp}$ and doubly integrated intensity (DIN), proportional to the EPR susceptibility $\chi_{EPR}$. (Here it is worthwhile mentioning that analysis of temperature dependences of EPR intensities provides data only on the spin contribution to susceptibility ($\chi_{EPR} \sim \chi_{CW}$ and $\chi_{Pauli}$) excluding its core ($\chi_{core}$) and orbital ($\chi_{orb}$) contributions. Electron spin-lattice relaxation times $T_{1e}$ were estimated using progressive microwave-power saturation technique. Precise calibration of *g*-factor was done by simultaneous recording of samples under study with a capillary tube containing small amount of 1 mM/l water solution of TEMPOL ($g_{iso}$ = 2.0059 ± 0.0001). Spectra' processing, calculations of parameters and spectra' simulation were done using Bruker WIN-EPR and OriginLab software. NMR measurements were performed using a Tecmag APOLLO pulse solid state NMR spectrometer, an Oxford Instruments 360/89 superconducting magnet ($B_0$ = 8.0196 T) and an Oxford Instruments CF1250 cryostat. $^{13}$C NMR spectra ($f_0$ = 85.85 MHz) were recorded using Fourier transformation of the phase cycled Hahn echo. Measurements of the $^{13}$C nuclear spin-lattice relaxation time $T_{1n}$ were done using progressive saturation pulse sequence. $^{13}$C NMR measurements on astralen were done in the temperature range 77(±0.5) K ≤ $T$ ≤ 290(±0.5) K, while the spectrum of the graphite powder was measured at ambient temperature. $^{13}$C chemical shifts $\sigma$ are given relative to tetramethylsilane (TMS).

## 3. RESULTS
### 3.1 STRUCTURE OF ASTRALEN NANOPARTICLES

HRTEM provides direct evidence that astralen particles have a polyhedral multi-shell structure with a hollow interior (Fig. 1, a-b). The average diameter of astralen particles is ~ 40 nm. The polyhedra consist of flat graphitic faces with average size of

~ 15 nm connected by defective edge regions (Fig. 1, c-d). The faces comprise a stacking of 20-50 planar graphene sheets with an inter-sheet distance of ~ 0.340 nm.

Fig. 2a shows the XRD patterns of the graphite reference and astralen powder samples. The former includes narrow peaks of (002), (100), (101) and (004) crystal planes of bulk graphite. The position of the (002) peak corresponds to the conventional graphite mean inter-layer spacing $d_{002}$ of 0.335 nm. In the astralen powder sample, the position of the (002) peak is shifted to the low angle direction corresponding to the mean inter-layer spacing $d_{002}$ of 0.340 nm between the graphitic shells in the polyhedral particles. This value is completely consistent with that obtained by the HRTEM observations.

Fig. 2b demonstrates representative Raman spectra of the graphite reference and astralen powder samples. Both spectra show two broad Raman bands around 1340 nm (*D*-band) and 1580 cm$^{-1}$ (*G*-band). Our graphite reference and astralen powder samples demonstrated ratios between peak amplitudes $I_D/I_G$ ~ 0.12 and ~ 0.29, correspondingly. Ratios between integrated intensities (areas) of *D*- and *G*-bands, fitted with the Lorentzian function are of 0.26 and 0.44, respectively.

### 3.2 EPR

Room temperature (RT, $T$ = 300 K) EPR spectrum of astralen particles, recorded within the maximal scan width of 1 T at incident microwave power $P$ = 20 mW and 100 kHz modulation amplitude of 1 mT, consists of two clearly distinguished components: a very broad EPR signal with the line width $\Delta H_{pp}$ exceeding 1 T and an asymmetric narrow ($\Delta H_{pp}$ of the order of several mT) signal centered close to $g$ = 2.00 (Fig. 3a, black solid line). The sharp low field ($g$ ~ 4.3) EPR line in these spectra (see Fig. 3a) belongs to $Fe^{3+}$ ions originating from the capillary tube' glass and is used as an external intensity reference. All spectral components (both broad and narrow ones) are found to be practically independent on the ambient pressure and show no visible changes when the sample was pumped out to the vacuum level better than $2 \times 10^{-5}$ mbar at both RT and $T = 600$ K. EPR measurements at variable temperatures were carried out on the sample sealed under vacuum. The black solid line in Figure 3b shows RT spectrum of the $g = 2.00$ EPR signal recorded within the reduced scan width of 40 mT at conditions of higher spectral resolution (modulation amplitude of 0.1 mT). This RT spectrum may be successfully simulated as superposition of two Lorentzian lines[14] (Fig. 3b, red dashed line) of different line widths, g-factors and intensities: $\Delta H_{pp1} = 0.67 \pm 0.05$ mT, $g_1 = 2.0034 \pm 0.0002$ and $\Delta H_{pp2} = 3.6 \pm 0.3$ mT, $g_2 = 2.007 \pm 0.001$, intensities ratio $DIN_2/DIN_1 \sim 13$ - see Fig. 3b, green dotted and blue dash-dotted line.

Since possible extension of the electron spin-lattice relaxation times $T_{1e}$ on decreasing temperature may affect the true values of the EPR lines' intensities (due to the saturation effect) the saturation curves were measured at selected temperatures. No EPR signals' saturation was observed within the temperature range under study - see Fig. 4a. Saturation curves at all temperatures coincide with each other and are perfectly linear. Therefore $T_{1e}$ values may be estimated as being shorter than $T_{1e} = 5 \times 10^{-8}$ s reported for the crystalline 2,2-diphenyl-1-picrylhydrazyl (DPPH) sample[15] that reveals saturation behavior at $P > 50$ mW ( Fig. 4a, closed circles).

EPR signals associated with astralen demonstrate unusual temperature evolution. It was found that the intensities of the broad and two $g = 2.00$ signals, unlike EPR signals of the most paramagnetic species, do not obey the Curie law. Figs. 3a and 3c obviously demonstrate very weak changes of astralen EPR signals' intensity on decreasing temperature from $T = 300$ K down to $T = 4$ K whereas the EPR signal of paramagnetic $Fe^{3+}$ ions from the sample tube, being under the same conditions as the astralen sample, shows evident temperature dependence of its intensity - see corresponding low-field signals in Fig. 3a. Fig. 4b represents temperature dependences of the intensities of two $g = 2.00$ signals. These intensities were obtained by different methods: DIN by direct double numerical integration of the experimental spectra (narrow and broad Lorentzian lines together), S by simple calculation using experimental data for the narrow Lorentzian line intensity and the line width as $S = I_{pp} \times (\Delta H_{pp})^2$ and by spectra' simulation (narrow and broad Lorentzian lines separately). Double integration of the observable part of the very broad EPR lines with $\Delta H_{pp} > 1$ T (Fig. 3a), done after the subtraction of $Fe^{3+}$ ions' and narrow $g = 2.00$ signals, evidences that these intensities are temperature independent (though within the large experimental error, not shown) as well. Both $g$-factor and line width of the broad Lorentzian line slightly increase on temperature decrease and then jump-down at $T = 4$ K (Fig. 4, c–d). The $g$-factor of the narrow line insignificantly decreases from 2.0034 down to 2.0030 and its line broadens for a few hundredths of mT.

Amounts of the paramagnetic centers observed were obtained at RT by comparison of intensities of the $g = 2.00$ signals with the intensity of the radical-like signal in well purified nanodiamond sample with known amount of paramagnetic defects ($N_s = 6.3 \times 10^{19}$ spins g$^{-1}$).[16] Here it is worth mentioning that this astralen sample shows strong non-resonant microwave absorption, which leads to the reduction of the spectrometer sensitivity. Thus, all data on absolute amounts of paramagnetic centers should be considered as lower limits of corresponding values and bearing an evaluative character. The amount of paramagnetic centers is found to be $N_{s1} \sim 5 \times 10^{17}$ spins g$^{-1}$ for the narrow Lorentzian line, $N_{s2} \sim 7 \times 10^{18}$ spins g$^{-1}$ for the broad Lorentzian line and, for the very broad ($\Delta H_{pp} > 1$ T) line, $N_{sbr}$ exceeds $10^{22}$ spins g$^{-1}$ (just an observable part of this line was taken for the estimation).

EPR spectra of pure graphite powder (not shown) demonstrate the same general features as found in astralen samples, i.e. superposition of the very broad ($\Delta H_{pp}$ > 1 T) line and the narrow asymmetric line in the region of $g$ = 2.00 (like those in Fig. 3a). The asymmetric line demonstrates line shape typical for the axially symmetric $g$-tensor with $g_{\parallel}$ = 2.031 ± 0.005 and $g_{\perp}$ = 2.007 ± 0.001 (at $T$ = 300 K). The intensity of this line also does not obey the Curie law but notably increases on decreasing temperature down to 40 K. Below 40 K the intensity starts declining.

### 3.3 NMR

$^{13}$C NMR spectra of the astralen powder and reference graphite samples recorded at $T$ = 290 K are shown in Fig. 5a. Both spectra show similar asymmetric lines due to chemical shielding anisotropy. Thus we conclude that the spectrum of astralen is characteristic of $sp^2$-graphite-like carbons. However, the astralen spectrum is found to be a little broader in comparison with that of graphite and its center of gravity is shifted by ~15 ppm to the lower frequency. The above broadening is presumably caused by a number of non-equivalent carbon sites in the multi-shell structure of astralen, say, due to stacking faults of the layers and defects.

$^{13}$C nuclear magnetization recovery, obtained using progressive saturation pulse sequence, in both graphite and astralen samples is well fitted by a stretched exponential

$$M(t) = M_0 \left\{ 1 - \exp\left[ -\left( \frac{t}{T_{1n}} \right)^{\alpha} \right] \right\}, \tag{1}$$

where $M_0$ is an equilibrium magnetization and $T_{1n}$ is $^{13}$C nuclear spin-lattice relaxation time. In graphite, $\alpha$ was found to be ~ 0.89, while in astralen the averaged value of $\alpha$ in the temperature range 77–290 K is 0.65. RT relaxation time $T_{1n}$ in our graphite sample is found to be $T_{1n}$ = 106 ± 8 s, which is quite close to that reported by Carver[17] in pristine graphite powder, $T_{1n}$ = 89 ± 10 s, and by Kume[18] in high oriented pyrolytic graphite (HOPG), $T_{1n}$ around 100 s. RT $T_{1n}$ value in astralen was found to be 152 ± 15 s.

Fig. 5b shows temperature dependence of $T_{1n}$ and $T_{1n} \times T$ values for astralen particles in the range from 77 to 290 K, which yields $T_{1n} \sim T^{-0.612}$.

## 4. DISCUSSION

Our HRTEM, XRD and Raman spectroscopy studies revealed a polyhedral multi-shell fullerene-like structure of astralen particles that consist of large flat $sp^2$ graphene faces connected by defective corner regions with presumably pentagon-like structure. The spacing of lattice fringes of 0.340 nm is larger than the (002) inter-plane distance in graphite (~ 0.335 nm) but smaller than the corresponding value in polyhedral carbon onions from HT annealed nanodiamonds (0.345 nm[9] and 0.353 nm[7]). The larger intersheet distance suggests a considerable reduction in the interlayer interaction compared to the case of bulk regular graphite. Fitting of the (002) peak of the XRD pattern of astralens (Fig. 2a) with the Lorentzian function gives an FWHM value of about 0.6°. It corresponds to the size of a coherent scattering region (CSR) of ~ 16 nm for the direction normal to the graphitic layers, calculated with the Scherrer formula. This value (i.e. 47 graphene sheets) is also in accord with the HRTEM observations. The broad and asymmetric (10) peak at ~ 45° is caused by the superposition of the (100) and (101) peaks that can be attributed to XRD scattering from a 2D lattice. These findings point out to the turbostratic structure and lack of 3D ordering in multi-shell astralen particles whose parallel graphitic sheets are randomly translated and rotated along the azimuth.[19] Fitting this peak by the superposition of two Lorentzian peaks (100) and (101) gives the FWHM values of 0.7° and 2.9°, for the (100) and (101) peaks, correspondingly. Thus, the CSR value calculated with the Scherrer formula in the lateral (100) direction (along the graphene plane) is ~ 15 nm. This value is consistent with the average size of the flat faces of polyhedral astralen particles revealed by HRTEM. Therefore, we suggest that these carbon nanoparticles have defect-free $sp^2$ flat faces and all of the defects condense at their polyhedral edges.

Analysis of Raman spectra of astralen samples (Fig. 2b) seems to be more controversial. The band at about 1580 cm$^{-1}$ ($G$-band) corresponds to the $E_{2g}$ mode in the graphite structure. Ideal single-crystalline graphite should show only the $G$-band in the spectrum range from 1200 to 1700 cm$^{-1}$.[20] In addition to the $G$-band, a so-called $D$-band appears at about 1340 cm$^{-1}$ for finite-size or disordered graphite-like samples (e.g. polycrystalline graphite, glassy carbon, etc.). The true origin of the $D$-band in graphite-like materials is still under discussion.[21,22] However, it was empirically shown that a modest amount of disorder and the resultant decrease in an in-plane domain size ($L_a$) of the graphitic $sp^2$ sheets could give rise to the $D$-band.[19,23] Accordingly, the relative intensity of the $D$- to $G$-bands ($I_D/I_G$) is widely used for a qualitative representation of $L_a$ in graphite-like samples in general and carbon onions, in particular,[5,7] using the empirical formula[23]

$$L_a \text{ (nm)} = 4.4 \times (I_D/I_G)^{-1}, \qquad (2)$$

where $I_D/I_G$ is ratio between peak amplitudes of $D$- and $G$-bands. Using Eq. (2) we estimate the in-plane size $L_a$ as 15 nm that is in accordance with the corresponding CSR value obtained by XRD and HRTEM observation.

However, recently [24-25] it was suggested that the $I_D/I_G$ ratio depends strongly on the excitation laser energy ($E_l$) used in the Raman experiment, while formula (2) was only valid when the experiment was done using the $\lambda_l$ = 514.5 nm ($E_l$ =2.41 eV) laser line. Studying diamond-like carbon films, heat treated at different temperatures and, thus, giving rise to nanographites with different $L_a$ values, Cançado et al. ,[25] revealed that $I_D/I_G$ is inversely proportional to the fourth power of $E_l$ as:

$$L_a \text{ (nm)} = (2.4 \times 10^{-10}) \lambda_l^4 (I_D/I_G)^{-1}, \qquad (3)$$

where $I_D/I_G$ is ratio between integrated intensities (areas) of *D*- and *G*-bands.

Determination of $L_a$ using Eq. (3) provides us with an unrealistic value of 88 nm that is more than twice larger than the average diameter of astralen particles observed by HRTEM. Adequacy of Eq. (3) has never been checked for different types of nanocarbon samples while the decrease in $I_D/I_G$ ratio with the laser energy as $E_l^4$ was independently confirmed only for ball-milled nanographite.[26] Although this issue is beyond the scope of this paper, we can conclude that correct determination of $L_a$ in various nanocarbon systems by Raman spectroscopy remains an open question so far and constitutes stuff for a future study.

Anyway, in-plane dimensions of the defect-free $sp^2$ flat faces in astralen particles are considerably larger than those for the best polyhedral carbon onions originated from HT treated nanodiamonds: 3.6 nm [9] and 7 nm.[7] On the other hand, it should be noted that giant polyhedral carbon particles with flat face sizes up to 100 nm were recently synthesized by laser ablation[27] and DC arc vaporization[28] of graphite. Unfortunately, electronic properties of these particles have not been reported.

Let us now consider the magnetic resonance data obtained for astralens in the view of the results and conclusions drawn from the thorough studies of quasi-spherical and polyhedral carbon onions originated from HTT nanodiamonds.[5,7–9]

Except for the very the broad EPR line with $\Delta H_{pp} > 1$ T, both relatively narrow signals with $g = 2.00$ revealed in our study have also been observed in EPR spectra of various HTT nanodiamonds. Thus, narrow ($\Delta H_{pp} \sim 0.7–0.9$ mT) Lorentzian lines with $g = 2.0020–2.0022$, which intensities obey the Curie law, have been found in quasi-spherical carbon onions obtained by "soft" thermal treatment.[5,9] Broad ($\Delta H_{pp} \sim 1–10.9$ mT, depending on the oxygen pressure) Lorentzian lines with $g = 2.0010 – 2.0014$, which intensities show features characteristic of Pauli or mixed Pauli-Curie paramagnetism, have been found in small polyhedral particles obtained by more intensive thermal treatment.[7,9] These relatively broad EPR signals were attributed to non-bonding π-electrons localized at the marginal regions of a nanographite sheet having a zigzag shape.[9] The Lorentzian shape observed for the $g = 2.00$ EPR lines is quite a natural line shape for de-aggregated conducting carbon nanoparticles where particle dimensions are smaller than the skin depth at ~ 9.4 GHz (which exceeds 500 nm even for metallic copper and gold).

In general, the $g = 2.00$ EPR signals found in astralen (Fig. 3c) look like signals of the aforementioned types and belong to some defects in carbon network. However, there are several noteworthy differences found for the signals in astralen in comparison with those observed in carbon onions produced by HT treatment of nanodiamonds:

(i) Both EPR signals in astralen do not depend on oxygen pressure whereas the broad EPR signals from nanodiamond originated particles were found to be very sensitive to oxygen pressure. Indeed, we found that reduction of oxygen pressure does not affect the line widths (within the corresponding experimental errors) for any EPR signals observed in astralen. Broadening of the EPR signals due to the spins localized at the edge states in an oxygen atmosphere has been studied in detail and is considered to be due to dipole-dipole and exchange interactions between the paramagnetic dioxygen molecules and the spins localized at edges and/or other defects.[9,29,30] It means that at ambient conditions oxygen molecules cannot penetrate through the graphene sheets of astralen particles. This finding is in accord with an absence of HRTEM observations of any structural voids or discontinuities in the graphene layers of astralens;

(ii)     intensities of all EPR signals (including the narrow and the very broad ones) do not depend on temperature within the very broad temperature range demonstrating the Pauli paramagnetism (Fig. 4b). Only the narrow $g = 2.00$ signal reveals a very weak Curie-like behavior on approaching $T = 4$ K - see hexagons in Fig. 4b;

(iii)    $g$-factor values measured for the broad and narrow Lorentzian lines in astralen are found to be higher than those for HT treated nanodiamonds: 2.0034 vs. 2.0022 and 2.007 vs. 2.0013 for the narrow and broad signals at RT, respectively.

The electronic properties of nanographite systems including their magnetic and conducting characteristics depend on sizes of defect-free domains in graphene sheets, $L_a$.[9] In our astralen particles the size of these defect-free domains coincides with the size of flat graphene faces. On the other hand, the $L_a$ is comparable (and even exceeds) the known size for nano-graphite particles exhibiting an abrupt transition (in the range of 5–15 nm) to the bulk graphite properties.[31]

The most unusual feature of the EPR properties observed in astralen is the Pauli-type behavior of EPR magnetic susceptibilities for all signals in the wide temperature range, especially in the low-temperature region 4–50 K (Fig. 4b). Such a behavior has never been observed in EPR studies done on multi-shell nanocarbon samples. Thus, nanodiamond-derived onion-like and polyhedral nanoparticles demonstrated the Curie-Weiss behavior below 100 K for both narrow and broad lines with the Curie-type contribution originating from localized $sp^3$ spins (narrow lines) and the enhanced Pauli-type contribution originating from non-binding π-electron states spins quasi-localized at the zigzag edges (broad lines).[7,8] It was shown that the spin paramagnetism associated with the edge state behaves as a temperature-dependent Curie-type or an enhanced Pauli-type temperature independent susceptibility according to the feature of the edge state.[9] The interaction of the edge state with the conduction band carriers creates the Pauli-type behavior and the structural disorder leads to the Curie-type localized nature for the edge-state spins.

Following Osipov et al.,[9] we speculate that the narrow and broad Lorentzian EPR signals observed in our samples are associated with localized and quasi-localized spins, respectively, though the corresponding defects in astralens and the onions produced by HT treatment of nanodiamonds can be of different origin. Indeed, in astralen both the quasi-localized and the localized spins are associated with the corner-condensed defects and interact with numerous conduction electrons which belong to the same graphene layer of the astralen particle. Accordingly, unlike the case of small multi-layered particles produced by HT treatment of nanodiamonds, where the treatment conditions determine the observed features of paramagnetic behavior for the edge-state spins, in astralen not only itinerant spins (the very broad line) but also quasi-localized spins (the broad $g$ = 2.00 line) show the same Pauli-type behavior down to 4 K. In our opinion, the unusual properties of the EPR active spins observed in astralen samples can be understood within the framework of the well-developed theory of electron resonance of localized electron spins in metals.[32,33] Following this theory the main factor determining all electron magnetic resonance features is the ratio of magnetic susceptibilities of conduction electrons $\chi_s$ and localized paramagnetic centers $\chi_d$: $\chi_r = \dfrac{\chi_d}{\chi_s}$. In the cases of $\chi_r \ll 1$ the conduction electrons' properties will predominate and EPR signals of localized paramagnetic centers will behave in the same manner as signals of conduction electrons spin resonance (CESR), i.e. their intensities will demonstrate the Pauli-type paramagnetism. When $\chi_r > 1$ the properties of EPR signal will be determined by properties of localized paramagnetic centers, that is the Curie-type paramagnetism. Thus, the aforementioned assumption on the high value of ratio between the numbers of delocalized π–electrons and localized and quasi-localized spins in astralen system exhaustively explains the unusual temperature dependences of the EPR signals observed in astralen. Several experimental observations support this assumption. First of all, the aforementioned strong non-resonant microwave absorption found (which leads to the Q-value reduction) is a characteristic feature for all conduction graphite samples. Another argument in favor of the presence of a large amount of conduction electrons may be the very broad ($\Delta H_{pp} > 1$ T) EPR signals that have been observed in both astralen and conducting graphite samples. It was carefully proven that in our experiments this line distinctly stood out against a background (i.e. signals due to cavity, quartz Dewar insert, glass capillary tube).

It is well known that in most metals, CESR lines may be so broad that CESR is not observable.[32,33] Observation of broad EPR signals in natural graphite powder was reported in Reference [34]. Quite broad ($\Delta H_{pp} > 0.1$ T) CESR signals were observed in $Rb_3C_{60}$ system[35] and doped chalcogenide nanotubes.[36] The origin of the large widths in both metals and other systems lies outside the scope of the present paper. Thus, very broad CESR lines are not uncommon ones for carbon containing and nano-sized systems. Moreover, the same very broad EPR line we observed in our reference graphite sample for which intensity of the signal of localized defects also shows the Pauli-type paramagnetism above 40 K.

On the other hand the alternative origin of that very broad EPR line observed in both astralen and powdered graphite can not be absolutely ruled out. In general this EPR line may occur due to some abundance of ferromagnetic impurities with the ferromagnetic transition point $T_C$ well above RT. However, typical ferromagnetic impurities in nano-carbon samples demonstrate much narrower EPR line.[30,37] Moreover, it is hard supposing the same impurities (responsible for the same very broad EPR lines) appear in both astralen and commercial pure graphite. Taking into account all aforementioned speculations the attribution of the very broad EPR line to CESR sounds quite reasonable.

Let us estimate the number of different spins per each astralen particle. The bottle density of astralen was found to be $2.2 \pm 0.1$ g cm$^{-3}$.[10] The average diameter of an astralen particle is 40 nm which gives the weight of a single astralen particle $7.38 \times 10^{-17}$ g. Thus each astralen particle contains (totally, in all graphene layers) ca. 40 localized spins, ca. 500 quasi-localized spins and above $10^6$ spins responsible for the CESR signal.

As one can clearly see from Fig. 5 in the Reference [7], in nanodiamonds-originated multi-shell nanoparticles both broad and narrow EPR signals demonstrate the Curie-type paramagnetism below 100 K. The same behavior was found in the nanodiamond sample heat treated for 2 min, where the ratio between the narrow signal (localized spins) and the broad one (quasi-localized π-electrons) was found to be ca. 20, i.e. comparable with that found in astralen. This means that the quasi-localized π-electrons only, even supposing they obeys the predominant Pauli-type behavior, which was found to be correct just above 100 K,[7,9] can not suppress the

Curie-type behavior. On the other hand, the presence of more than a million of conduction (delocalized) π–electrons per each astralen nanoparticle perfectly complies with the condition $\chi_r \ll 1$ and dictates the distinctly pronounced Pauli-type behavior for intensities of all EPR signals observed.

Our hypothesis on fulfilling the $\chi_r \ll 1$ condition in astralen assumes the availability of strong exchange interaction between conduction electron spins and localized spins, responsible for the broad and narrow $g = 2.00$ EPR signals. It was shown[32,33] that this exchange interaction leads to the shift of the resonance field for localized spins due to the effective hyperfine field induced by conduction electrons. This effect may explain low-field shift of resonance fields (positive $g$-shifts mentioned above) of both EPR lines attributed to localized and quasi-localized spins in comparison with the same values found in carbon onions produced by HTT of nanodiamonds.[5,7–9] The latter samples demonstrate the mixed Curie-Pauli paramagnetism and evidently higher $\chi_r$ values that reduces corresponding exchange $g$-shifts.

The temperature-dependent $^{13}$C NMR spin-lattice relaxation data in astralen supplies additional arguments in favor of the aforementioned hypothesis. The non-exponential nuclear spin-lattice relaxation in solids is usually caused by a distribution of relaxation times within the sample. Such a relaxation may be well analyzed using the stretched exponential curve (Eq. 1), where $\alpha$ is a fitting parameter. The presence of different relaxation environments of the carbon atoms in the polyhedral astralen onions (due to the non-equivalence of the carbon atoms mentioned above) may be the reason for the observed relaxation behavior. Alternatively, such behavior may be attributed to the interaction of nuclear spins with localized paramagnetic centers. Spin-lattice relaxation via paramagnetic centers usually leads to a noticeable reduction in $T_{1n}$, However, such a reduction is not observed in our experiment which complies with quite low density of localized paramagnetic centers found in astralen sample (~ 5 × $10^{17}$ spins $g^{-1}$ as estimated by EPR).

$^{13}$C NMR relaxation measurements on pristine graphite powder done at 1.3–4.2 K and room temperature, made by Carver[17], led to $T_{1n} \sim T^{-0.7}$ dependence. The same measurements on astralen in the range of 77 to 290 K yield $T_{1n} \sim T^{-0.612}$ (Fig. 5b). This finding shows that the slope value in the temperature dependence of $T_1$ falls between the metallic behavior, $T_{1n} \sim T^{-1}$ (Korringa relation), and the semiconductor behavior, $T_{1n} \sim T^{-0.5}$.

Finally, both EPR and NMR experiments justify the following model of the electron system in astralen. Each large multi-shell carbon polyhedron contains: (a) relatively small (ca. 40) number of localized spins; (b) quasi-localized spins (ca. 500); (c) spins of delocalized (conduction) π–electrons (above $10^6$). Both localized and quasi-localized spins are condensed near the corners of the astralen polyhedra.

## 5. CONCLUSIVE REMARKS

We report unusual electronic and magnetic properties of the astralen nanoparticles and discuss them in terms of the hypothesis that each astralen nanoparticle constitutes a closed network of delocalized π-electrons. This feature, in turn, is attributed to the multi-shell polyhedral structure of astralen and in particular to the fact that the polyhedra consist of large (~ 15 nm) defect-free $sp^2$ flat faces while a limited number of defects condense at the polyhedral edges.

## ACKNOWLEDGEMENTS

This work was supported, in part, by the Israeli Ministry of Immigrant Absorption and the New Energy Development Organization of Japan (NEDO, grant # 04IT4). The authors acknowledge Dr. L. Zeiri for her help with Raman measurements.

**Figure Captions**

Fig. 1. Representative HRTEM images of astralen nanoparticles.

Fig. 2. (Color online) XRD patterns a) and Raman spectra b) of the reference graphite (black dash-dotted line) and astralen (blue solid line) powder samples. The profiles are shifted along the vertical axis.

Fig. 3. (Color online) EPR spectra of astralen particles: a) general view at RT (black line) and $T$ = 4 K (red dashed line) recorded at the same instrumental conditions, $\nu$ = 9.465 GHz; b) asymmetric $g$ = 2.00 EPR signal, black solid line - experimental spectrum (RT, $\nu$ = 9.462 GHz), red dashed line - simulated spectrum, best least square fit using superposition of two Lorentzian signals: narrow signal (green dotted line) and broad one (blue dash-dotted line); c) $g$ = 2.00 EPR signal recorded at different temperatures: 4 K (black line), 100 K (red dashed line), 200 K (green dotted line), and RT (blue dash-dotted line). Spectra were recorded at the same instrumental conditions.

Fig. 4. (Color online) a) Microwave saturation dependences for the narrow signal at selected temperatures: circles - $T$ = 4 K, triangles up - $T$ = 40 K, diamonds - $T$ = 160 K, stars - $T$ = 300 K, red dashed line - best least square linear fit. Blue closed circles - saturation curve for the DPPH sample, $T$ = 300 K; b) Temperature dependences of the integral intensities of $g$ = 2.00 EPR signals': stars - experimental DIN obtained by numerical integration of both $g$ = 2.00 signals; hexagons - intensity $S$ of the narrow Lorentzian signal calculated as

$S = I_{pp} \times (\Delta H_{pp})^2$; intensities of the narrow (diamonds) and broad (pentagons) Lorentzian signals obtained by simulation; circles - intensity $S$ of paramagnetic $Fe^{3+}$ ions signal. Signal intensities are normalized to the same values at $T = 300$ K; Temperature dependences of $g$-factors c) and line widths d) for the broad (red closed circles) and narrow (open black circles) Lorentzian $g = 2.00$ EPR signals.

Fig. 5. (Color online) a) $^{13}$C NMR spectra of the astralen powder (blue dotted line) and reference graphite (black solid line) samples recorded at $T = 290$ K; b) temperature dependences of $^{13}$C nuclear spin-lattice relaxation time $T_{1n}$ (black solid circles) and $T_{1n} \times T$ values (inset, open red circles) in the astralen sample.

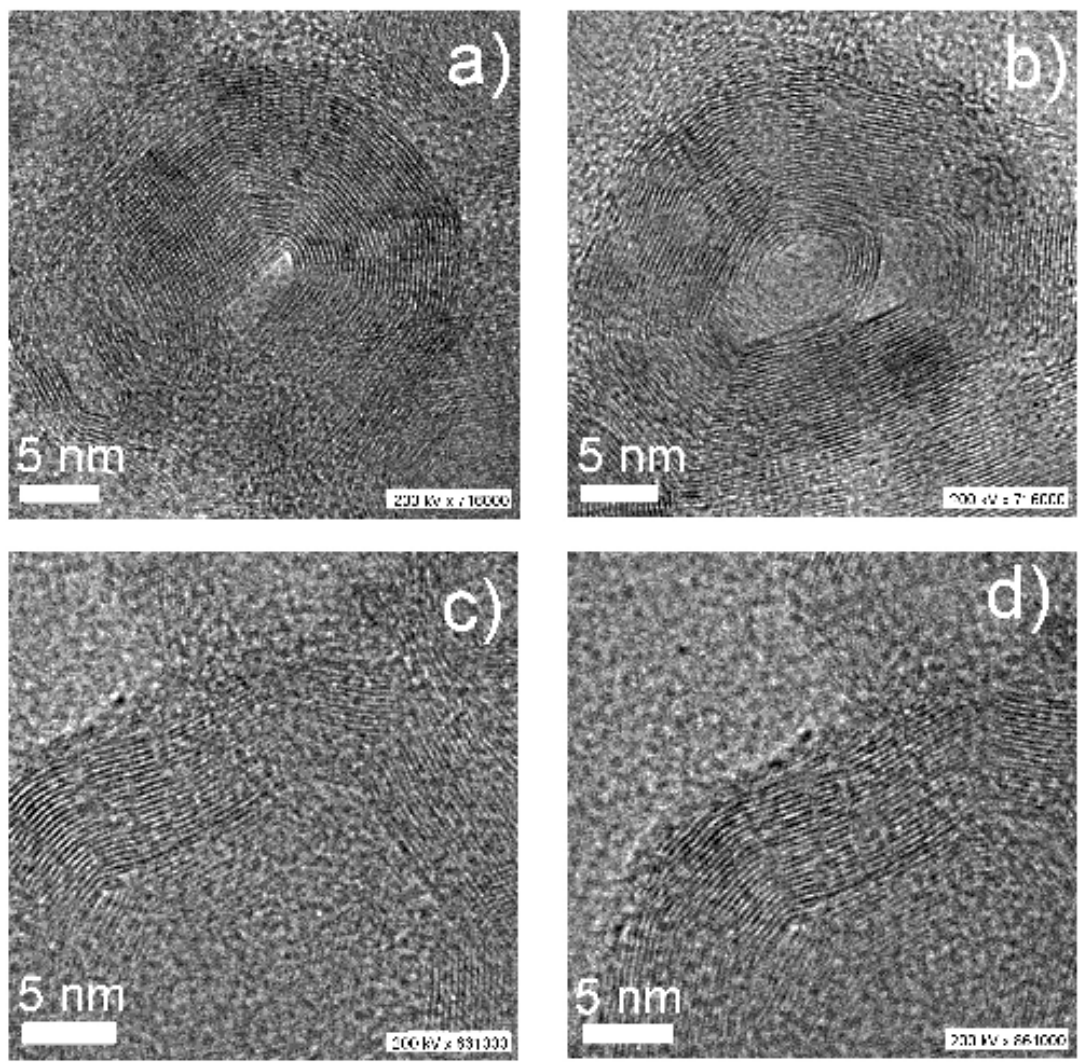

Fig. 1.

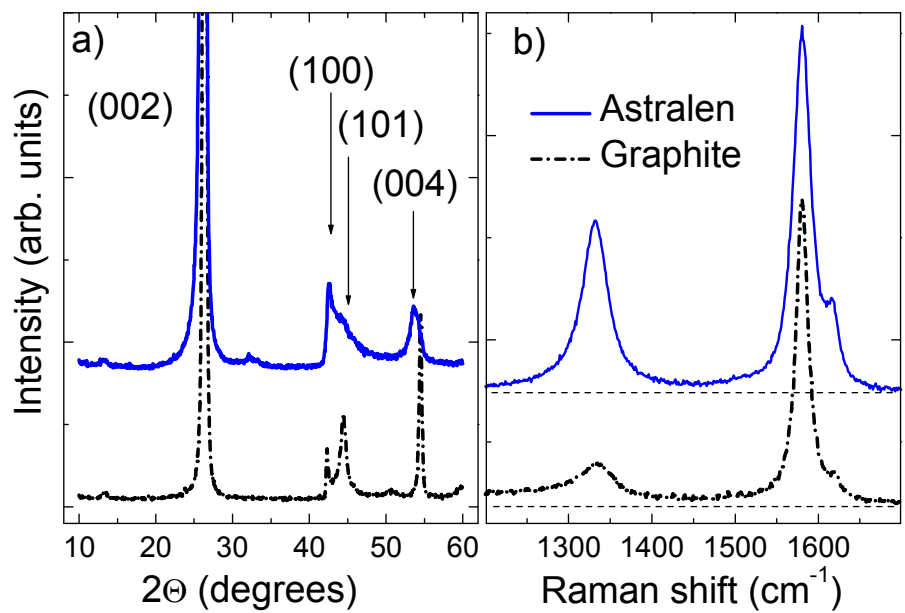

Fig. 2.

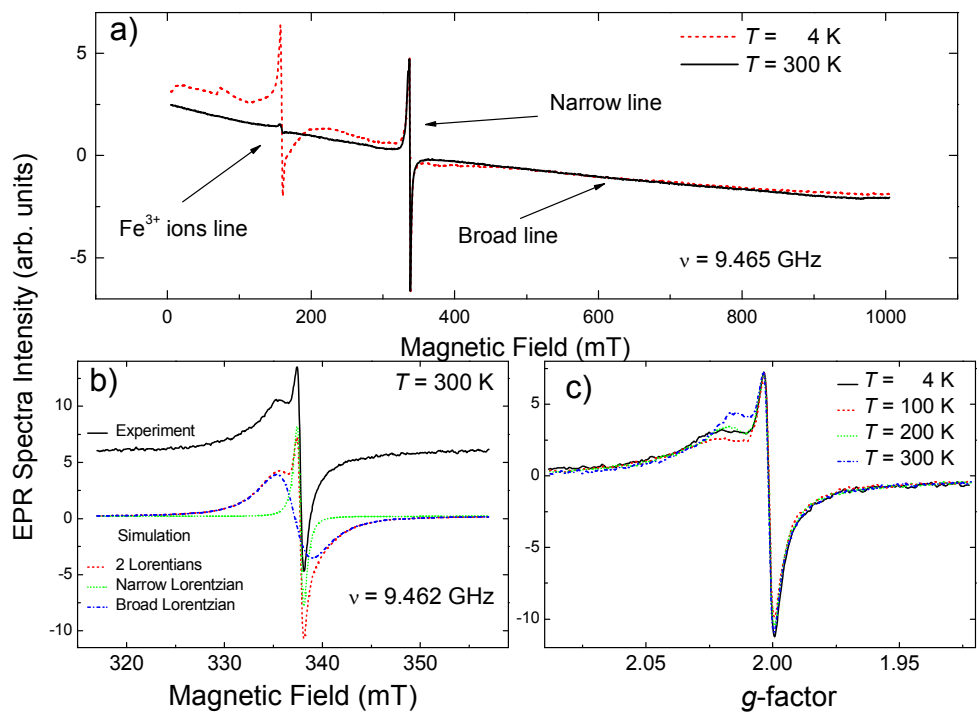

Fig. 3.

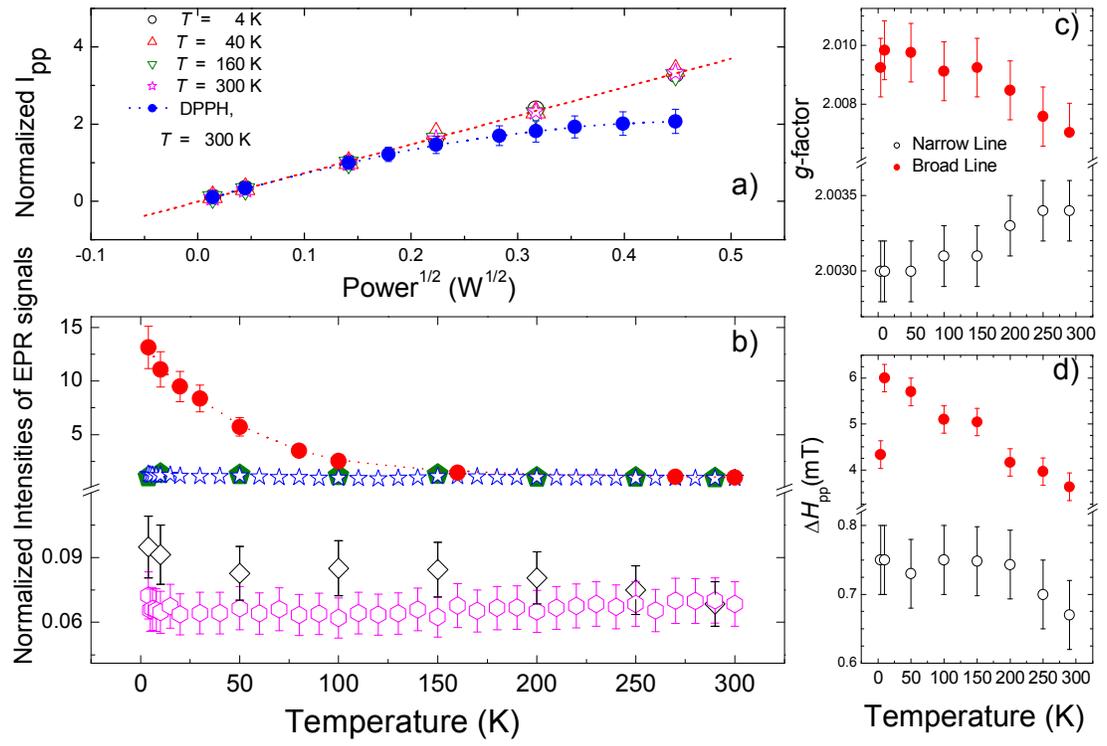

Fig. 4

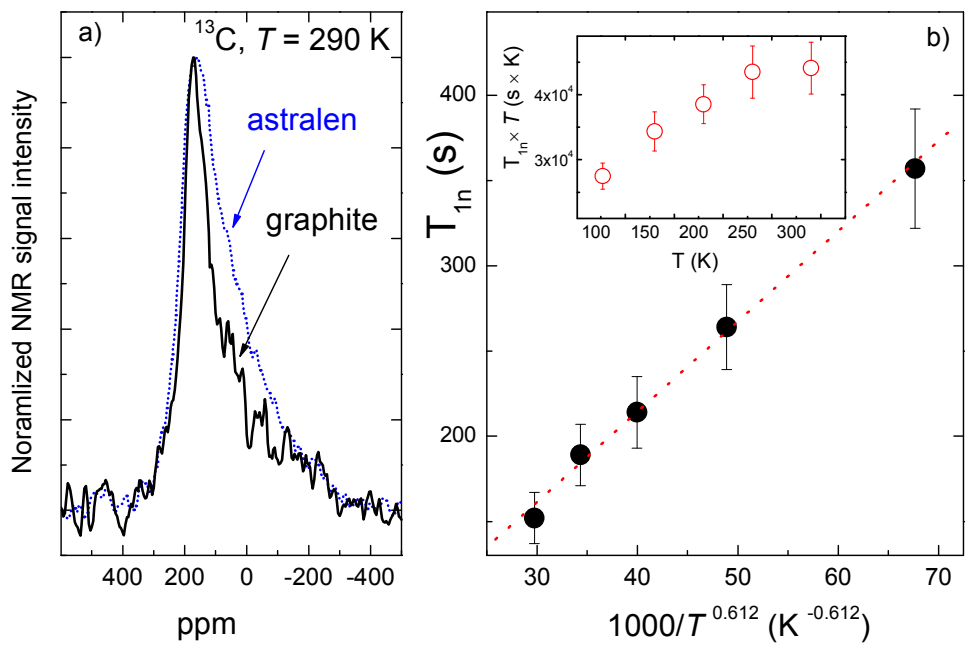

Fig. 5.